\documentclass{PoS}

\usepackage{cite}

\title{Resonances in $W_LW_L$, $Z_LZ_L$ and $hh$ scattering from dispersive analysis of the non-linear Electroweak+Higgs Effective Theory.}

\ShortTitle{Resonances in $W_LW_L$, $Z_LZ_L$ and $hh$ scattering}

\author{\speaker{Antonio Dobado}\thanks{{\tt dobado@fis.ucm.es}; Contribution PoS(EPS-HEP2015)173. }}
\author{Rafael L. Delgado and Felipe J. Llanes-Estrada~\thanks{On leave at the Institute of Nuclear Theory, Univ. of Washington}\\
        Dept. F\'{\i}sica Te\'orica I, Fac. CC. F\'{\i}sicas, Plaza de las Ciencias 1, 28040 Madrid, Spain}
\abstract{If new resonances of the electroweak symmetry breaking sector (longitudinal-gauge and Higgs) bosons are found in the 1-3 TeV region, the right tool to assess their properties and confront experimental data in a largely model-independent yet simple manner is Unitarized Effective Theory. 
Its ingredients are: 1) custodial symmetry and the Equivalence Theorem, that allow to approximate $W_L$ and $Z_L$ by an isospin-triplet of Goldstone bosons $\omega^a$ in the 1~TeV region. 2) The effective coupling of a generic, approximately massless scalar-isoscalar $h$ to those Goldstone bosons, and the chiral Lagrangian describing them, valid up to about 3~TeV. 3) The Inverse Amplitude or other unitarization techniques that allow to extend the reach of perturbation theory to the first resonance in each partial wave. \\
We highlight some of the parameter space that can give rise to 2~TeV resonances, for example a simultaneous scalar-isoscalar and a vector-isovector ones (motivated by the ATLAS excess) and also the potential importance of coupled-channel dynamics between $hh$ and $\omega\omega$.
}
\FullConference{The European Physical Society Conference on High Energy Physics\\
		22--29 July 2015\\
		Vienna, Austria}

\begin{document}
%%%%%%%%%%%%%%%%%%%%%%%%%%%%%%%%%%%%%%%%%%%%%%%%%%%%%%%%%%%%%%%%%%%%%%%%%
The LHC is exploring the 1-3 TeV energy region in the Electroweak Symmetry Breaking Sector (and perhaps finding new resonances there) which motivates developing and adapting theoretical methods to treat any such resonances.
Since the LHC shows that the low-energy, few-hundred GeV limit of the theory, contains only the $W$ and $Z$ bosons and the new Higgs-like $h$ boson, an economic description of that lowest energy part is to formulate an effective Lagrangian for these particles alone. A minimal extension thereof, ``Unitarized Effective Theory'' then allows to cover the two-body resonances that may appear below $4\pi v\sim 3$ TeV. 

Under the aegis of the Equivalence Theorem~\cite{Cornwall:1974km}, scattering amplitudes for longitudinal gauge bosons $W_LW_L$ and $Z_LZ_L$ can be substituted by much simpler Goldstone-boson ones $\omega^a$, with order $(M_W^2/s)$ corrections . It is then consistent to neglect $M_W^2$, $M_Z^2$ and $m_h^2$, all around (100 GeV)$^2$, against the s-scale (1 TeV)$^2$. 
The effective Lagrangian~\cite{Buchalla:2015wfa,Alonso:2014wta,Kilian:2014zja,Delgado:2013loa,Gripaios:2015qya} then has seven parameters, $a$, $b$, $a_4$, $a_5$, $g$, $d$, and $e$,
\begin{eqnarray} \label{bosonLagrangian} {\cal L}
& = & \frac{1}{2}\left[1 +2 a \frac{h}{v} +b\left(\frac{h}{v}\right)^2\right]
\partial_\mu \omega^i \partial^\mu
\omega^j\left(\delta_{ij}+\frac{\omega^i\omega^j}{v^2}\right) \nonumber
+\frac{1}{2}\partial_\mu h \partial^\mu h \nonumber  \\
 & + & \frac{4 a_4}{v^4}\partial_\mu \omega^i\partial_\nu \omega^i\partial^\mu
 \omega^j\partial^\nu \omega^j +
\frac{4 a_5}{v^4}\partial_\mu \omega^i\partial^\mu \omega^i\partial_\nu
\omega^j\partial^\nu \omega^j  +\frac{g}{v^4} (\partial_\mu h \partial^\mu h )^2
 \nonumber   \\
 & + & \frac{2 d}{v^4} \partial_\mu h\partial^\mu h\partial_\nu \omega^i
 \partial^\nu\omega^i
+\frac{2 e}{v^4} \partial_\mu h\partial^\nu h\partial^\mu \omega^i
\partial_\nu\omega^i \ .
\end{eqnarray}

The partial-wave amplitudes in NLO effective theory have the generic form
\begin{equation} \label{pertamplitude}
A_{IJ}^{(0)}(s) + A_{IJ}^{(1)}(s) = 
K s + \left( B(\mu)+D\log\frac{s}{\mu^2}+E\log\frac{-s}{\mu^2}\right) s^2 
\end{equation}
where $K$ and $E$ are related by perturbative unitarity for physical $s$, ${\rm Im} A^{(1)}_{IJ} = | A^{(0)}_{IJ}|^2$, and $B(\mu)$ contains the NLO low-energy constants that absorb one-loop divergences and ensure order by order renormalizability. $K$ is proportional to $(1-a^2)$ (in elastic $\omega\omega$ channels) and to $(a^2-b)$ (in inelastic $\omega\omega\to hh$ channels). Thus, in the Standard Model, where $a=1=b^2$, the tree-level amplitude does not grow as $s$; any parameter separation signals strong interactions and makes the SM a fine tuned (though renormalizable) parameter choice. In the complex $s$-plane the amplitudes present both left and right cuts due to intermediate particle-loops in $s$, $t$ and $u$ channels as usual. 

However, as is well known from hadron physics, if resonances are present the energy range of validity of the effective theory is much reduced, possibly 
%beging reduced 
to the area near threshold (where the equivalence theorem does not apply anyway). This is understood as a failure of exact unitarity, that is only satisfied up to NNLO corrections in perturbation theory. To guarantee the correct unitarity and analyticity properties, one resorts to dispersion relations whose numbers not directly obtainable from data (left cut and subtraction constants) are fixed in perturbation theory, for example. The method is called ``Unitarized Effective Theory''~\cite{Truong:1988zp,Dobado:1989gr}.

It is convenient to define an auxiliary right-cut carrying function $g(s)$
as well as to split the NLO amplitudes $A^{(1)}$ in a left-cut carrying and a right-cut carrying parts as follows,
\begin{eqnarray}
\nonumber
& g(s)\equiv\frac{1}{\pi}\left(\frac{B(\mu)}{D+E}+\log\frac{-s}{\mu^2}\right)\ ; \\
& A_L(s)       \equiv  \left(\frac{B(\mu)}{D+E}+\log\frac{s}{\mu^2}\right) Ds^2\ ; 
\ \ \ \ \ \ \ \ \ \ 
A_R(s)       \equiv  \left(\frac{B(\mu)}{D+E}+\log\frac{-s}{\mu^2}\right) E s^2   \ .    
\end{eqnarray}
%\begin{eqnarray}
%\nonumber
%g(s)&\equiv& \frac{1}{\pi}\left(\frac{B(\mu)}{D+E}+\log\frac{-s}{\mu^2}\right)\ ; \\
%A_L(s)       \equiv  \left(\frac{B(\mu)}{D+E}+\log\frac{s}{\mu^2}\right) & D&s^2\ ; 
%\ \ \ \ \ \ \ \ \ \ 
%A_R(s)       \equiv  \left(\frac{B(\mu)}{D+E}+\log\frac{-s}{\mu^2}\right) E s^2   \ .    
%\end{eqnarray}
The split $B/(D+E)$ is designed so that all pieces are separately renormalizable and $\mu$ independent. 

Then one can construct (at least) three algebraic formulae from the perturbative partial wave amplitude that effect the unitarization. All three have the correct analytic structure, being derived from a dispersive approach, satisfy exact elastic unitarity, match to perturbation theory when expanded at low-energy, and allow for resonances in the complex plane; they are the Inverse Amplitude Method, a variant of the N/D method and an improved K-matrix method~\cite{Delgado:PRD},
\begin{eqnarray} \label{Atogether}
A^{IAM}(s) =  \frac{\left[A^{(0)}(s)\right]^2}{A^{(0)}(s)-A^{(1)}(s)} 
 = \frac{A^{(0)}(s)+A_L(s)}{1-\frac{A_R(s)}{A^{(0)}(s)} -\left(\frac{A_L(s)}{A^{(0)}(s)}\right)^2 +g(s)A_L(s)};\nonumber\\
A^{N/D}(s) =    \frac{A^{(0)}(s) + A_L(s)}{1-\frac{A_R(s)}{A^{(0)}(s)} +\frac{1}{2}g(s)A_L(-s)};\ \  \ \ \ \ \ \ \ \ \ 
A^{IK} (s) =    \frac{A^{(0)}(s)+A_L(s)}{1-\frac{A_R(s)}{A^{(0)}(s)} +g(s)A_L(s)}. \nonumber
\end{eqnarray}

\begin{figure}
\includegraphics[width=0.48\textwidth]{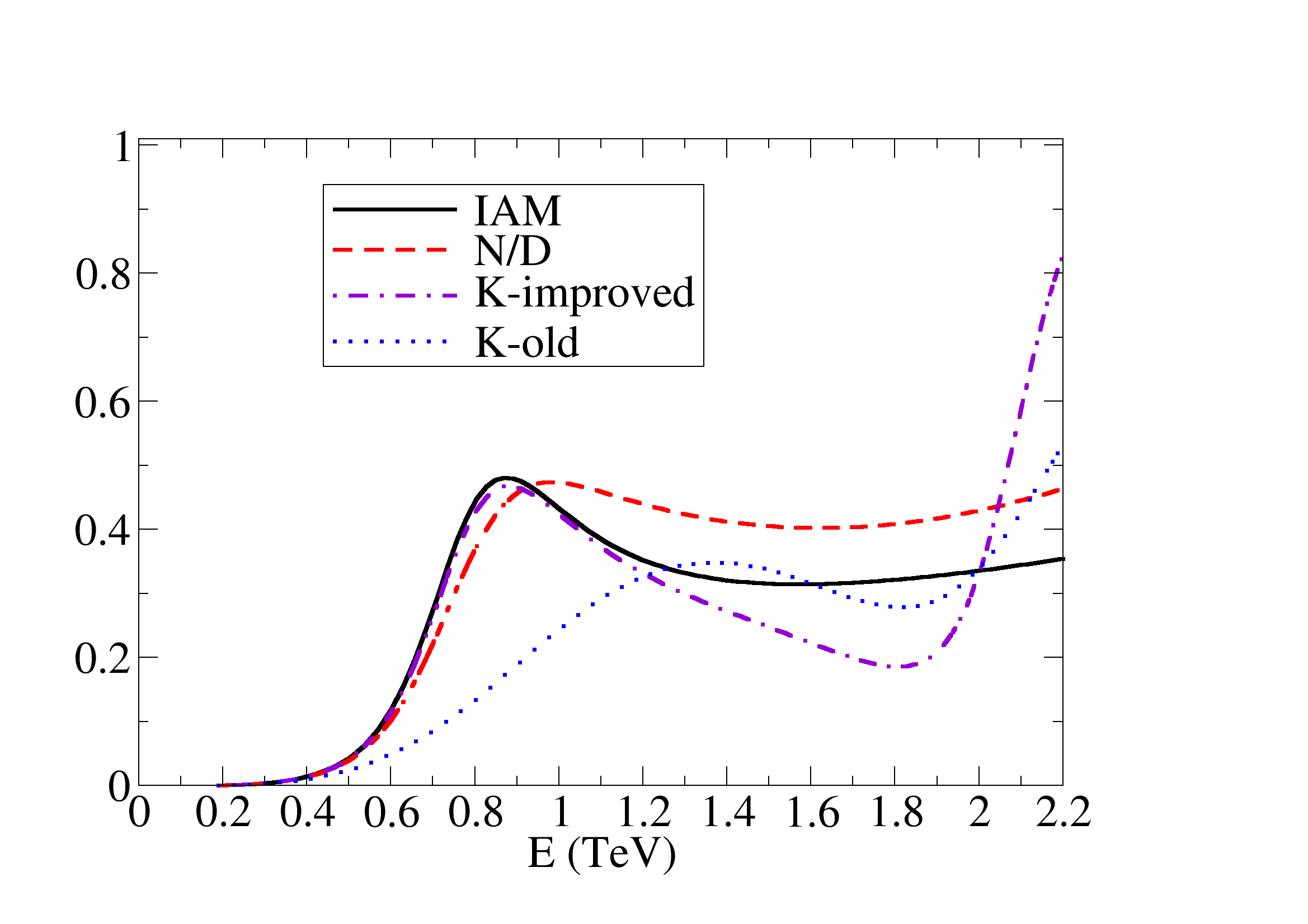}
\includegraphics[width=0.48\textwidth]{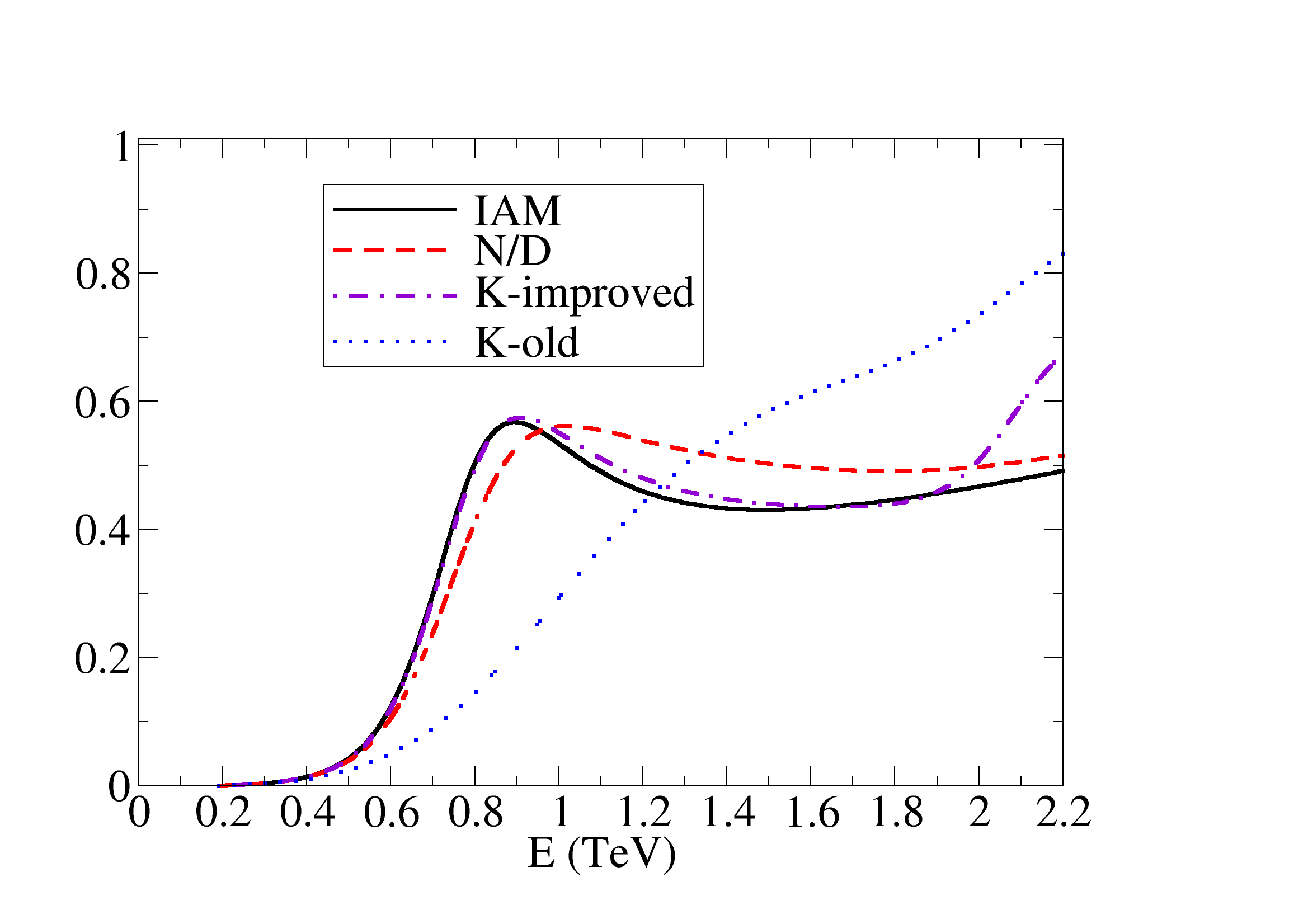} \\
\begin{minipage}{0.48\textwidth}
\includegraphics[width=0.95\textwidth]{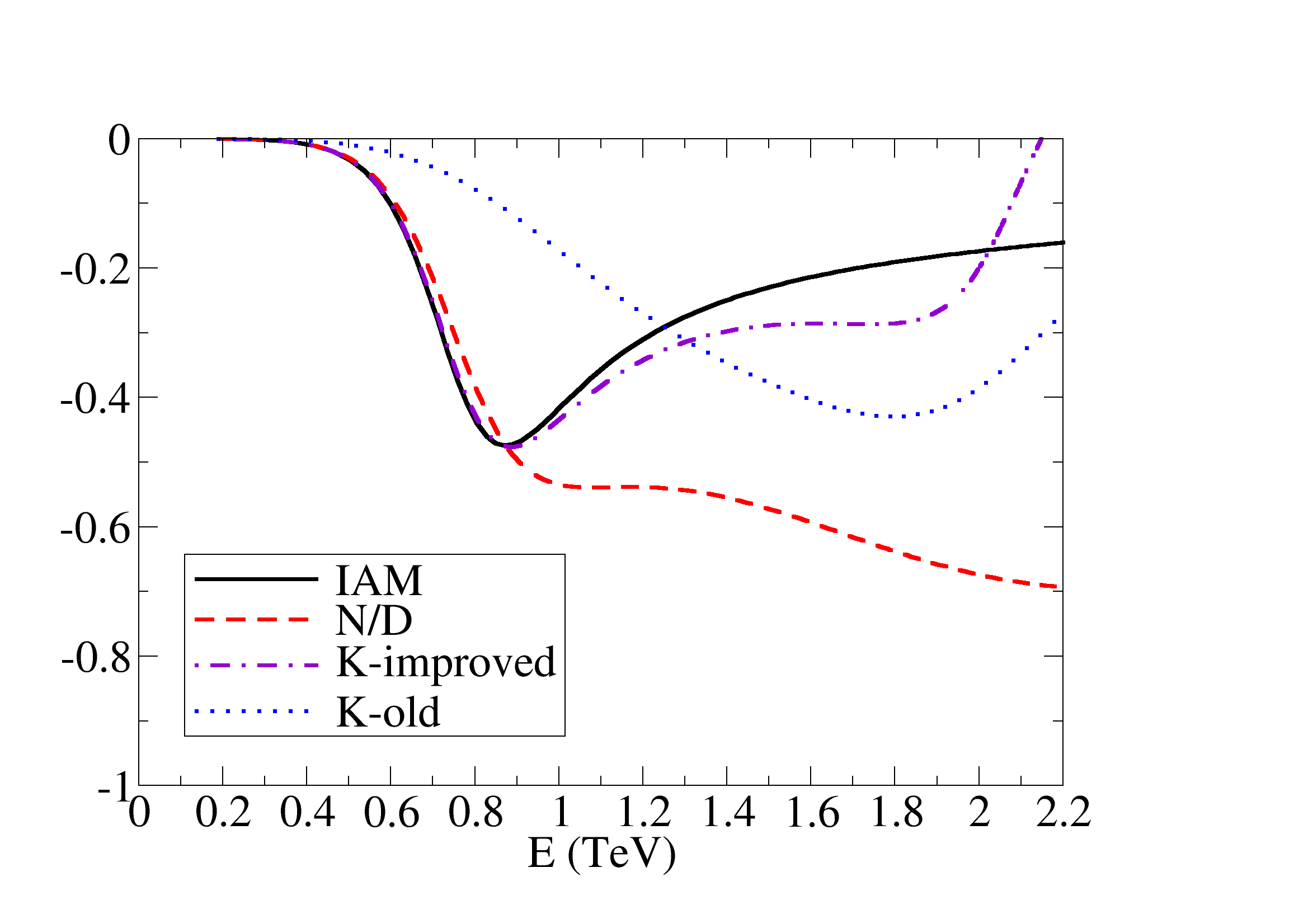}
\end{minipage}
\begin{minipage}{0.48\textwidth}
\caption{\label{fig:IAM} We compare the three unitarization methods
%of Eq.~(\ref{Atogether}) 
for the imaginary parts of the $IJ=00$ amplitudes.
Clockwise from top left, $\omega\omega$, $hh$ and cross-channel $\omega\omega\to hh$ (with LO parameters $a=0.88$ and $b=3$, $\mu=3$ TeV and the
NLO ones set to zero). A scalar resonance is visible in all, and the
unitarization methods with correct analytic properties closely agree (the disenting old-K matrix method is unitary but not analytic).}
\end{minipage}
\end{figure}

If the right logarithm dominates over the left one, $A_R\gg A_L$, all three unitarization methods yield the same resonances at approximately the same positions, since it is $A_L$ that marks the differences among them. This can be seen in figure~\ref{fig:IAM} where the NLO counterterms in $B(\mu)$ have been set to zero, so that it is the iteration of the LO amplitude that generates the dynamical pole.

\begin{figure}
\includegraphics[width=0.475\textwidth]{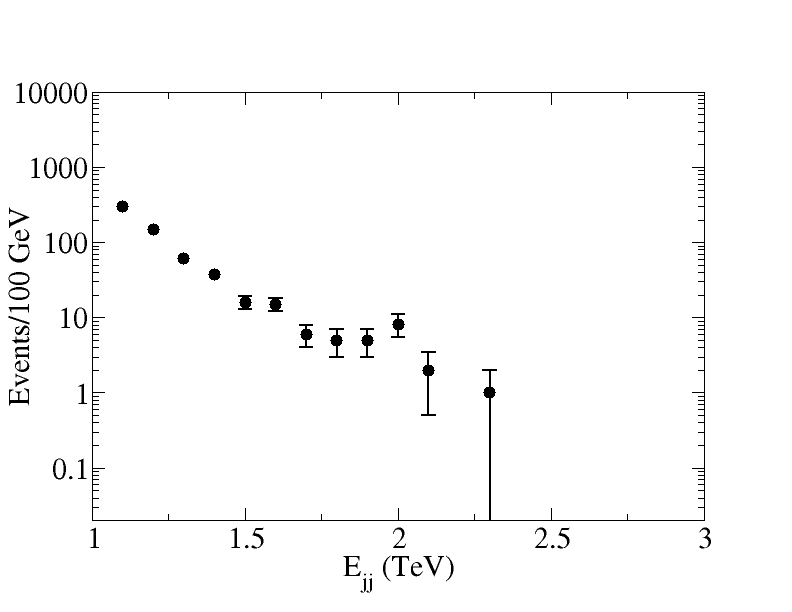}
\includegraphics[width=0.475\textwidth]{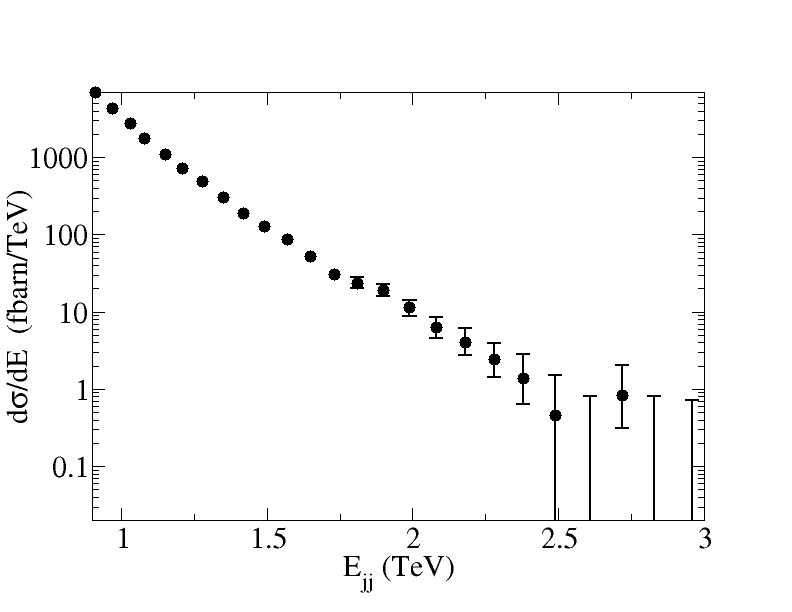}
\begin{minipage}{0.475\textwidth}
\centerline{\includegraphics[width=0.95\textwidth]{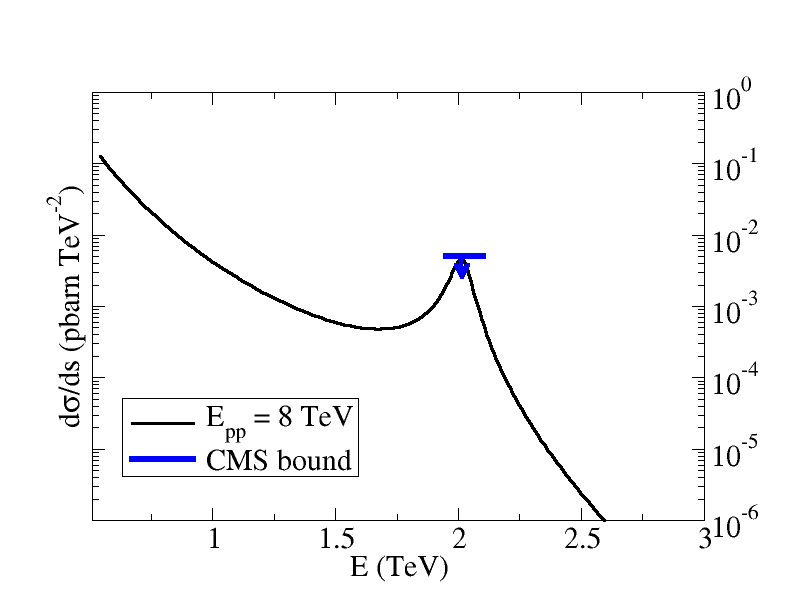}}
\end{minipage}
\begin{minipage}{0.475\textwidth}  
\caption{\label{fig:ATLASdata} Top: we replot the ATLAS\cite{Aad:2015owa} (left) and  CMS~\cite{Khachatryan:2014hpa} (right) data
for $WZ\to 2\ {\rm jet}$ in pp collisions at the LHC, that shows a slight excess at 2 TeV (same in the other isospin combinations $WW$ and $ZZ$, not shown) in ATLAS at 2 TeV, but not in CMS, ignoring the small excess at 1.8-1.9 TeV. In the bottom plot we give the tree-level production cross-section of $\omega\omega$ from~\cite{Dobado:2015hha} 
with a $IJ=11$ resonance in the final-state, for $a=0.9$, $b=a^2$, $a_4=7\times 10^{-4}$ (at $\mu=3$ TeV), together with the CMS upper bound on the cross-section.}
\end{minipage}
\end{figure}

Many different resonances can be described until enough low-energy measurements constrain all the parameters in the Lagrangian density of Eq.~(\ref{bosonLagrangian}). So we focuse now on giving them a mass of about 2 TeV as needed to address the ATLAS excess in a two-gauge boson spectrum replotted in figure~\ref{fig:ATLASdata}. 
The Barcelona group~\cite{Arnan:2015csa} and us~\cite{Dobado:2015hha}  have recently discussed that an isotensor resonance, as well as a pair of an isoscalar and an isovector resonance, all in the lowest partial wave, could feed all channels where the ATLAS excess is seen. However, typical cross-sections as reported again in fig.~\ref{fig:ATLASdata} are small, in agreement with earlier CMS bounds. It looks odd that ATLAS finds a larger production than excluded by CMS (which is compatible with the theoretical computations yielding moderate cross-sections), and we hope that the LHC II run will help clarify whether we face a statistical fluctuation.

A broad QCD-$\sigma$-like scalar-isoscalar pole has been broadly discussed, but the tighter bounds $a\in (0.88,1.3)$ inferred from $hWW$ measurements~\cite{ATLAS:2014yka} start constraining it, and it cannot explain an excess in the (charged) $WZ$ channel (\emph{if both bosons are well identified}), so it requires a simultaneous isovector resonance to exist. 

\begin{figure}
\includegraphics[width=0.475\textwidth]{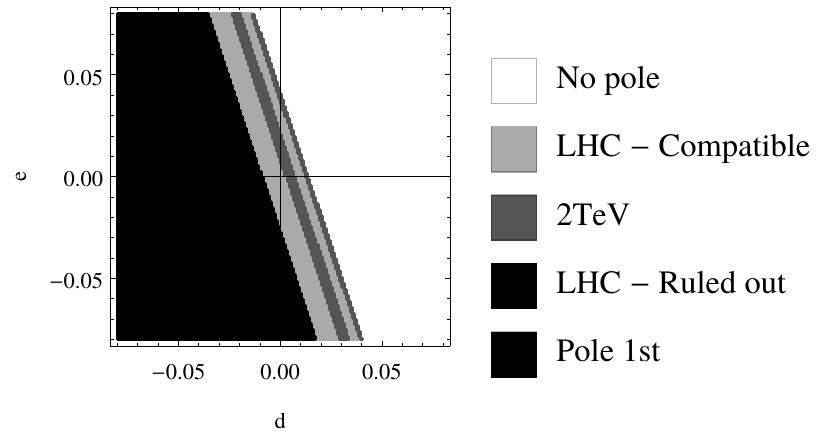}
\includegraphics[width=0.475\textwidth]{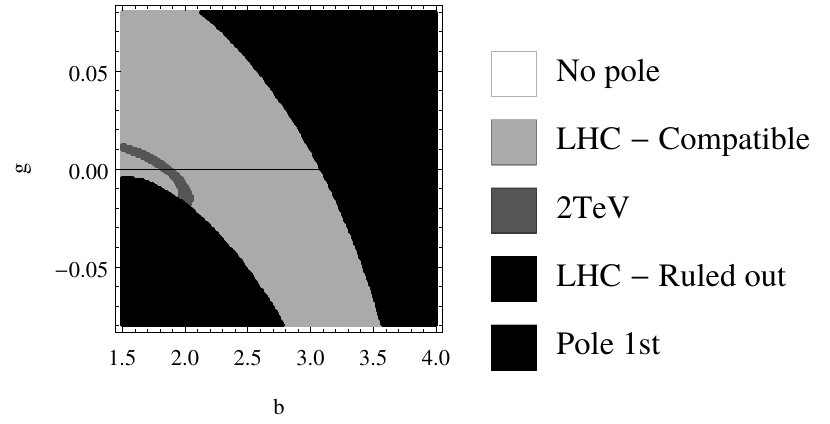}
\begin{minipage}{0.475\textwidth}
\centerline{\includegraphics[width=0.95\textwidth]{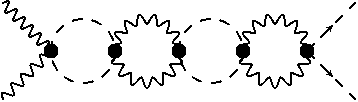}}
\end{minipage}
\begin{minipage}{0.475\textwidth}
\caption{\label{fig:pinballres} The dynamics of the pure Higgs sector can feed into the $\omega\omega$ spectrum, but also there can be resonant coupling between the two. Clockwise from top left:  parameter map for the NLO channel-coupling terms $d$ and $e$ from Eq.~(\protect\ref{bosonLagrangian}); and likewise for the $hhhh$ parameter $g$ with the LO coupling $b$, both maps showing the regions where $\omega\omega$ scattering poles are induced.
Finally, resonant diagram between the $hh$ and $\omega\omega$ channels resummed by the IAM and other coupled-channel methods.}
\end{minipage}
\end{figure}

A less trodden on possibility is that such pole be generated by dynamics resonating between the $\omega\omega$ and $hh$ coupled channels~\cite{Delgado:2015kxa}, see figure~\ref{fig:pinballres}; this pole can certainly occur at 2 TeV.
Figure~\ref{fig:pinballres} also shows, on the top right plot, how the pure $hh\to hh$ dynamics can feed into the $\omega\omega\to \omega\omega$ through the $b$ coupling. In our effective Lagrangian in Eq.~(\ref{bosonLagrangian}) we have only included one (derivative) interaction term of the pure Higgs sector, as we have not explored that dynamics in detail: this term with precoefficient $g$ is the only one necessary to achieve renormalizability up to NLO, but others may exist. Our published parameter maps extend the $a_4$-$a_5$ ones originally put forward by the Barcelona~\cite{Espriu} group to a complete study of the seven-parameter Lagrangian density.

\begin{figure}
\includegraphics[width=0.48\textwidth]{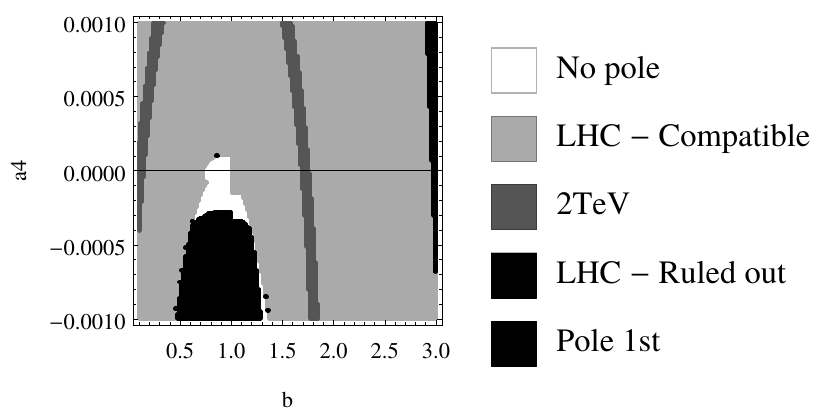}
\includegraphics[width=0.48\textwidth]{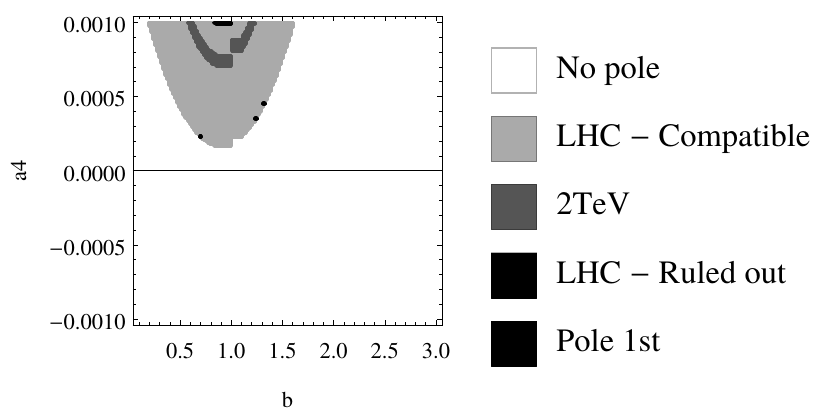} \\
\begin{minipage}{0.48\textwidth}
\includegraphics[width=0.95\textwidth]{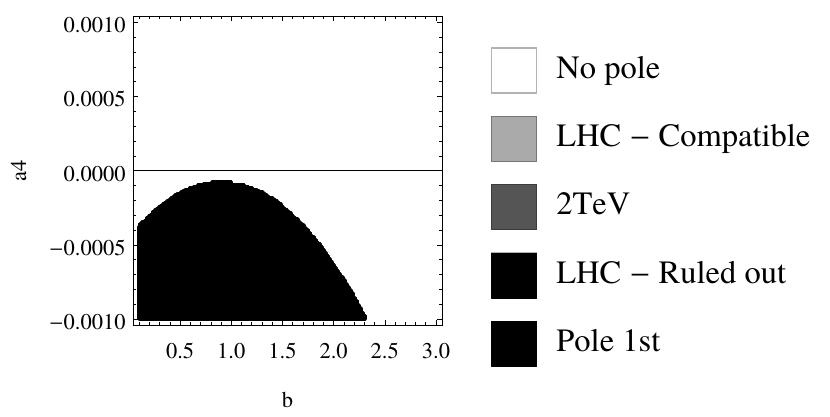} \\
\end{minipage}
\begin{minipage}{0.48\textwidth}
\caption{\label{fig:ba4} Searches for $\omega\omega$ scattering poles in $b-a_4$ parameter-space  for $a=0.95<1$. Clockwise from top left, $IJ=00$, $11$, $20$. The latter is not resonant, except for a pole in the 1st Riemann sheet (which excludes that swath of parameter space, in black). The first two show acceptable resonances in the 2nd Riemann sheet, also at 2 TeV (light gray). }
\end{minipage}
\end{figure}

In figure~\ref{fig:ba4} we seek poles in the complex $s$ plane
as function of the LO- $b$ and NLO-$a_4$ parameters for fixed $a=0.95$. 
Because $a<1$, $(1-a^2)>0$, and the isoscalar wave $A_0^0(s) = \frac{1}{16 \pi v^2} (1-a^2) s$  is attractive at LO while the isotensor one,
$A_2^0(s)  =  -\frac{1}{32 \pi v^2} (1-a^2)s$ is repulsive.

This is visible in the bottom plot of the figure: the isotensor channel presents no pole for positive or nearly positive $a_4$, and for negative $a_4$ (dark region) there is a pole in the first Riemann sheet, meaning that the isotensor interaction is then unacceptably repulsive, violating causality, so that part of the parameter space needs to be cut off (presumably no fundamental underlying theory can be matched to the effective theory with those parameters, or else the IAM must fail there).

However, the top plots of figure~\ref{fig:ba4} include thin, light-gray bands where an $IJ=00$ (left) or $IJ=11$ (right) pole is in the second Riemann sheet, with mass about 2 TeV. In fact, there are two spots in this parameter space, corresponding to about $a_4=0.0013$ and $b$ slightly smaller and slightly bigger than 1 respectively, where \emph{both} isoscalar and isovector poles are present near 2 TeV. That means that all of the $W_LW_L$, $W_LZ_L$ and $Z_LZ_L$ channels can be simultaneously fed.

In conclusion, if the LHC discovers new resonances coupled to the Electroweak Symmetry Breaking Sector of the Standard Model up to 3 TeV, Unitarized Effective Theory is currently positioned to describe that data
and map it to the few parameters of universal effective Lagrangians built on the non-linear sigma model. The necessary amplitudes can be given in simple analytical and algebraically closed form, as long as $s\gg m_W^2,m_Z^2,m_h^2$.

%%%%%%%%%%%%%%%%%%%%%%%%%%%%%%%%%%%%%%%%%%%%%%%%%%%%%%%%%%%%%%%%%%%
\section*{Acknowledgements}
%%%%%%%%%%%%%%%%%%%%%%%%%%%%%%%%%%%%%%%%%%%%%%%%%%%%%%%%%%%%%%%%%%%
We thank J. J. Sanz Cillero, F.-K. Guo and  D. Espriu for warm discussions. 
ADG thanks the organizers of the EPSHEP 2015 conference in Vienna for their excellent organization and inspiring working atmosphere
and the CERN TH-Unit for its hospitality, and FJLE likewise the Institute of Nuclear Theory of the Univ. of Washington and DOE support.
Work partially supported by Spanish Excellence Network on Hadronic Physics FIS2014-57026-REDT, and grants No. UCM:910309, MINECO:FPA2014-53375-C2-1-P, MINECO:BES-2012-056054 (R. L. D.).

%%%%%%%%%%%%%%%%%%%%%%%%%%%%%%%%%%%%%%%%%%%%%%%%%%%%%%%%%%%%%%%%%%%%%%%%%

%%%%%%%%%%%%%%%%%%%%%%%%%%%%%%%%%%%%%%%%%%%%%%%%%%%%%%%%%%%%%%%%%%%
\end{document}